\documentclass[useAMS,usenatbib,aas_macros]{mn2e} 
\usepackage {graphicx,subfigure,times,aas_macros}

\newcommand{\se}[1]{\S\ref{sec:#1}}

\newcommand{\Fig}[1]{Figure~\ref{fig:#1}}

\newcommand{\tab}[1]{Table~\ref{tab:#1}}
\newcommand{\be}{\begin{equation}}
\newcommand{\ee}{\end{equation}}
\newcommand{\bea}{\begin{eqnarray}}
\newcommand{\eea}{\end{eqnarray}}

\newcommand{\msun}{{\rm M}_\odot}
\newcommand{\Msun}{M_\odot}

\newcommand{\ifm}[1]{\relax\ifmmode#1\else$\mathsurround=0pt #1$\fi}
\newcommand{\kms}{\ifmmode\,{\rm km}\,{\rm s}^{-1}\else km$\,$s$^{-1}$\fi}

\newcommand{\kpc}{\,{\rm kpc}}

\newcommand{\ltsima}{$\; \buildrel < \over \sim \;$}
\newcommand{\lsim}{\lower.5ex\hbox{\ltsima}}
\newcommand{\gtsima}{$\; \buildrel > \over \sim \;$}
\newcommand{\gsim}{\lower.5ex\hbox{\gtsima}}

\def\la{\langle}

\def\cms{\,{\rm cm}^{-2}}

\def\Mv{M_{\rm v}}
\def\Rv{R_{\rm v}}

\def\Ms{M_*}

\def\re{r_{0.5}}
\def\T{\rm{Tri}}

\usepackage{color}

\begin{document} 

\large 

\title[elongated galaxies] {Formation of elongated galaxies with  low masses at high redshift}

\author[Ceverino et al.]
{Daniel Ceverino$^{1,2}$\thanks{E-mail: daniel.ceverino@uam.es}, 
Joel Primack$^3$, 
Avishai Dekel$^4$ \\
$^1$Centro de Astrobiolog{\'i}a (CSIC-INTA), Ctra de Torrej{\'o}n a Ajalvir, km 4, 28850 Torrej{\'o}n de Ardoz, Madrid, Spain \\
$^2$Astro-UAM, Universidad Autonoma de Madrid, Unidad Asociada CSIC \\
$^3$Department of Physics, University of California, Santa Cruz, CA, 95064, USA \\
$^4$Center for Astrophysics and Planetary Science, Racah Institute of Physics, The Hebrew University, Jerusalem 91904, Israel
}

\date{}

\pagerange{\pageref{firstpage}--\pageref{lastpage}} \pubyear{0000}

\maketitle

\label{firstpage}

\begin{abstract}

We report the identification of elongated (triaxial or prolate) galaxies in cosmological simulations at $z\simeq2$.
These are preferentially low-mass galaxies ($\Ms \le 10^{9.5} \ \Msun$), residing in dark-matter (DM) haloes with strongly elongated inner parts, a  common feature of high-redshift DM haloes in the $\Lambda$CDM cosmology.
Feedback 
slows formation of stars at the centres of these halos, so that a dominant and prolate DM distribution gives rise to galaxies elongated along the  DM major axis.
As galaxies grow in stellar mass, stars dominate the total mass within the galaxy half-mass radius, making stars and DM rounder and more oblate.
A large population of elongated galaxies produces a very asymmetric distribution of projected axis ratios, as observed in high-z galaxy surveys.
This indicates that the majority of the galaxies at high redshifts are not discs or spheroids but rather galaxies with elongated morphologies. 

\end{abstract}


\section{Introduction}
\label{sec:intro}

The intrinsic, three dimensional (3D) shapes of today's galaxies can be roughly described  as discs or spheroids, 
or a combination of the two.
These shapes are characterized by having no preferential long direction. 
Examples of galaxies elongated along a preferential direction (prolate or triaxial) are rare at $z=0$ \citep{Padilla08, Weijmans14}.
They are usually unrelaxed systems, such as ongoing mergers.
However, at high redshifts, $z=1-4$,
we may witness
the rise of the galaxy structures that we see today at the expense of other structures that may be more common during those early and violent times.

Observations trying to constrain the intrinsic shapes of the stellar components of high-z galaxies are scarce but they agree that
the distribution of projected axis ratios of high-z samples at $z=1.5-4$ is inconsistent with a population of randomly oriented disc galaxies \citep{Ravindranath06, Law12,Yuma12}. 
After some modeling, \citet{Law12} concluded that the intrinsic shapes are strongly triaxial. This implies that a large population of high-z galaxies are elongated along a preferential direction. 

\citet{vdWel14} looked at the mass and redshift dependence of the projected axis ratios  using a large sample of star-forming galaxies at $0<z<2.5$ from CANDELS+3D-HST and SDSS. 
They found that the fraction of intrinsically elongated galaxies increases toward higher redshifts and lower masses. 
They concluded that the majority of the star-forming galaxies with stellar masses of 
$\Ms = 10^9$ to $10^{9.5} \ \msun$ are elongated at $z \ge1$. 
At lower redshifts, galaxies with similar masses are mainly oblate, disc-like systems. 
It seems that most low-mass galaxies have not yet formed a regularly rotating stellar disc at 
$z \gsim 1$. 
This introduces an interesting theoretical challenge.
In principle, these galaxies are gas-rich and gas tends to settle in rotationally supported discs, if the angular momentum is conserved \citep{FallEfstathiou80, Blumenthal86, MoMaoWhite98, Bullock01}.
At the same time, high-mass galaxies tend  to be oblate systems even at high-z.
The observations thus 
suggest that protogalaxies
may develop an early 
 prolate shape and 
 then become oblate as they grow in mass.

Prolateness or triaxiality are common properties of DM haloes in N-body-only simulations \citep[][and references therein]{Jing02, Allgood06,Bett07,Maccio07,Maccio08,Schneider12}.
Haloes at a given mass scale are more prolate at earlier times,
and at a given redshift more massive halos are more elongated.
For example, small haloes with virial masses around $\Mv\simeq10^{11} \ \msun$ at redshift $z=2$ are as prolate as today's galaxy clusters.
Individual haloes are more prolate at earlier times, when haloes are fed by narrow DM filaments, including mergers, rather than isotropically, as described in \cite{VeraCiro11}. 
The progenitors of MW-size haloes are fairly prolate at redshift $z=2$
and they are increasingly more elongated at smaller radii \citep{Allgood06} because their inner shells collapsed earlier.

The shape of the inner DM halo could influence the shape of the central galaxy \citep{DekelShlosman83}.
If a triaxial halo dominates the inner gravitational potential, the inner galaxy feels a non-axisymmetric, non-rotating potential.
Under these conditions, stellar orbits elongated along the major axis of the triaxial halo are common within some characteristic radius of the system \citep[][ \S3.3.1 and 3.4.1]{BinneyTremaine}.
Therefore, elongated galaxies are expected within the $\Lambda$CDM paradigm. 

The 
fact that more massive dark matter 
haloes tend to be more prolate should be reconciled with the observation that
more massive galaxies, presumably living in more massive haloes, are more disky \citep{vdWel14}.
The missing piece in the picture of elongated galaxies is the counter-effect of baryons on the shape of the inner halo and its corresponding gravitational potential.
N-body+gasdynamics simulations of galaxy formation could provide this missing link.
\citet{Zemp12} compared the shape of  $z=2$ dark matter haloes in adaptive mesh refinement (AMR)  simulations with and without cooling, star formation and feedback. The addition of galaxy formation processes led to rounder haloes, especially in their inner parts, where baryons dominate the total mass.
However, these simulations had an excess of baryons 
compared to estimates based on abundance matching \citep{Moster13, Behroozi13} such that the simulated galaxies were always oblate, and the effect of baryons on the dark-matter halo was possibly overestimated.  

The Eris simulation of a galaxy with $M_v = 7.9 \times 10^{11} \ \msun$ at $z=0$ \citep{Guedes11} shows more realistic baryonic fractions at $z=0$ and 
it shows the formation of a pseudobulge at $z=4$ triggered by minor mergers that induced non-axisymmetric instabilities \citep{Guedes13}. The mass in this early pseudobulge was about $10^9 \Msun$, within the mass range of elongated galaxies.
Indeed, its stellar surface density map at $z=4$ looks marginally triaxial. However, there is no information about the DM mass and shapes in the inner regions.
 \cite{NihaoII} find 
 elongated halo shapes in N-body-only and hydro simulations at $z=0$ when the virial mass $M_{\rm vir} \lsim 10^{11} \msun$, while  
for higher $M_{\rm vir}$ the hydro simulations show substantially rounder halo shapes. 
However, they do not discuss the stellar shapes of their simulated galaxies or extend their analysis to higher redshifts, when haloes with low stellar-to-halo mass ratios are more elongated \citep{Allgood06}.
Thus, previous simulations 
have not revealed elongated galaxies as observed at $z \gsim 1$.

In this paper, we report the formation of elongated galaxies in cosmological hydrodynamic simulations.
We aim at understanding why elongated galaxies form preferentially in low-mass haloes at high redshifts.
We also address whether the gas distribution is also elongated and whether the simulations can
reproduce the typical shapes observed in the distribution of projected axis ratios in high-z galaxy surveys.
In section \se{runs} we describe the sample of simulations. \se{rise} describes the formation of elongated galaxies within prolate haloes and \se{fall} discusses their transformation into disk or oblate galaxies due to a high stellar central density. \se{compObs} shows the distribution of projected axis ratios and \se{conclusion} summarizes the results and discusses them.

\section{Simulations}
\label{sec:runs}

The zoom-in simulations used in this paper were drawn from a larger dataset, the  \textsc{Vela} sample, which contains 35 haloes with masses between $2 \times 10^{11} \ \msun$ and $2 \times 10^{12} \ \msun$ at $z=1$.
We selected the haloes that contain a galaxy with a stellar mass lower than $10^{9.5} \ \msun$ at $z=2$.
These are the best candidates for hosting elongated galaxies, according to observations \citep{vdWel14}.
The properties at $z=2$
of the five haloes that meet this criterion are described in \tab{1}. Their typical virial mass is $\Mv\simeq 10^{11} \ \msun$ and each hosts a galaxy with a stellar mass of $\Ms \simeq 10^9 \ \msun$, $0.6-1.2\%$ of the total virial mass.
This $\Ms/\Mv$ ratio is within a factor 2-3 of the estimates with abundance matching techniques, which are highly uncertain at this redshift and mass range \citep{Behroozi13}. 

The simulations were performed with the  \textsc{ART} code
\citep{Kravtsov97,Kravtsov03}, which accurately follows the evolution of a
gravitating N-body system and the Eulerian gas dynamics using an AMR approach.
Beyond gravity and hydrodynamics, the code incorporates 
many of the physical processes relevant for galaxy formation.  
These processes, representing subgrid 
physics, include gas cooling by atomic hydrogen and helium, metal and molecular 
hydrogen cooling, photoionization heating by a constant cosmological UV background with partial 
self-shielding, star formation and feedback, as described in 
\citet{Ceverino09}, \citet{CDB}, and \citet{Ceverino14}. 
 In addition to thermal-energy feedback, the simulations discussed here use radiative feedback.
This model adds a non-thermal pressure, radiation pressure, to the total gas pressure in regions where ionizing photons from massive stars are produced and trapped. 
In the current implementation, named RadPre in \citet{Ceverino14}, radiation pressure is included in the cells (and their closest neighbors) that contain stellar particles younger than 5 Myr and whose gas column density exceeds $10^{21}\ \cms.$ 
 
 The initial conditions of these runs contain $6.4-11 \times 10^6$ dark matter particles 
 with a minimum mass of 
$8.3 \times 10^4 \ \msun$, while the particles representing stars that were formed in the simulation
have a minimum mass of $10^3 \ \msun$. 
The maximum spatial resolution is between 17-35 proper pc. More details can be found in \citet{Ceverino14} and \citet{Zolotov}.

\begin{table} 
\caption{Properties of the zoom-in simulations at $z=2$. Columns show the name of the run,  the virial radius, the virial mass, the stellar half-mass radius, the stellar mass and gas mass. Masses are in $\Msun$ and radii in kpc.}
 \begin{center} 
 \begin{tabular}{cccccc} \hline 
\multicolumn{2}{c} {Run } \ \ $\Rv$ & $\Mv$ & $\re$ & stellar mass & gas mass \\
\hline 
Vela05   & 44 & $0.7 \times 10^{11}$  & 1.8  & $0.7  \times 10^9$ & $0.5  \times 10^9$\\ 
Vela04   & 53 & $1.2 \times 10^{11}$  & 1.7  & $0.8  \times 10^9$ & $0.8  \times 10^9$\\ 
Vela02   & 55 & $1.3 \times 10^{11}$  & 1.9  & $1.6  \times 10^9$ & $1.2  \times 10^9$\\ 
Vela28   & 63 & $2.0 \times 10^{11}$  & 2.3  & $1.8  \times 10^9$ & $2.1  \times 10^9$\\ 
Vela01   & 58 & $1.6 \times 10^{11}$  & 0.9  & $2.0  \times 10^9$ & $1.5  \times 10^9$\\ 
 \end{tabular} 
 \end{center} 
\label{tab:1} 
 \end{table} 

\section{Formation of Elongated Galaxies}
\label{sec:rise}

\begin{figure*} 
\includegraphics[width =0.99 \textwidth]{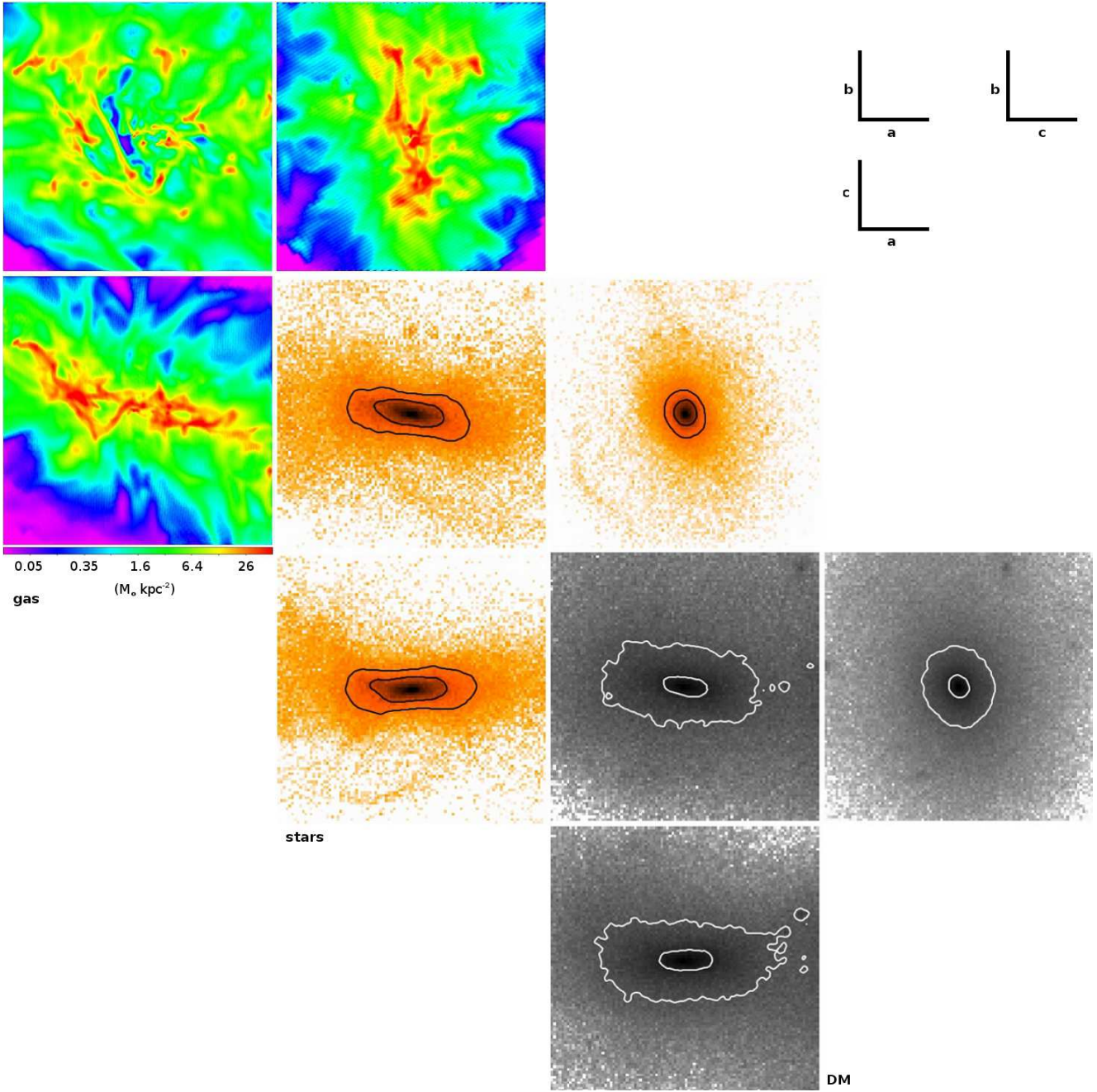}
\caption{Example of an elongated galaxy. Orthogonal views along the galaxy principal axes, showing gas, stars, and DM (from top to bottom) of Vela28 at $z=2.2$. The size of the images is 20 kpc. The contours of stellar surface density are defined as log($\Sigma/ (\Msun \kpc^{-2}$)  =\ 1, 1.5 
in the b-a and c-a planes and  as log($\Sigma/ (\Msun \kpc^{-2}$) =\ 1.5, 2 in the b-c plane. 
The stellar half-mass radius is $\re=2.1 \kpc$. 
The contours of DM surface density are set as log($\Sigma/ (\Msun \kpc^{-2}$) =\ 2, 2.5
along the minor and intermediate axes and as log($\Sigma/ (\Msun \kpc^{-2}$) =\ 2.2, 2.8 along the major axis.
The labels at the top-right corner mark the directions of the galaxy axes in each view.}
\label{fig:Vela28}
\end{figure*}

 \Fig{Vela28} shows three orthogonal views of the three components (gas, stars, and DM) of an elongated galaxy at $z=2.2$ (Vela28). The face-on view (b-a plane) was selected using the angular momentum of the cold ($T<10^4$ K) gas inside 5 kpc.
 The gas shows a very irregular and multiphase medium with clouds of relatively high column densities, log($\Sigma / \msun \ {\rm pc}^{-2} )=1.5-1.7$, and bubbles with very low densities. 
 The geometry is far from a uniform, thin disc, although the gas is flattened along a minor axis, $c$, coincident with the rotation axis of the gas. 
 Indeed, the gas rotates at ${\rm V}_{\rm rot}=80 \kms$, averaged over a 2-5 kpc shell. 
 The radial velocity dispersion measured at 200 pc scales is also high, $\sigma_{\rm r}=30 \kms$, typical of very perturbed, turbulent and thick discs. See \cite{Ceverino12} for more details.

The stellar distribution is very different from the thick gaseous disc. 
It is elongated along one direction, which defines the galaxy major axis $a$.
The 3D isodensity surface crossing the major axis around the half-mass radius is well fitted by an ellipsoid with axial lengths equal to $(a,b,c)=(2.2,0.76,0.67) \kpc$. 
The intermediate-to-major axis ratio is low, $b/a=0.35$, inconsistent with an axisymmetric disc. 
The shape is close to a prolate 
ellipsoid.

The dark matter also shows a prolate shape with similar properties at roughly the same radius, 
$(a,b,c)_{\rm DM}=(1.8,0.73,0.6) \kpc$ and $(b/a)_{\rm DM}=0.4$. 
The fact that the dark matter is also elongated in the same direction as the stellar component 
is relevant to the formation of such an elongated galaxy.

For a better understanding of the connection between the stellar and DM shapes, we perform a quantitative analysis of the intrinsic shapes of the stellar and DM distributions.
We select the major progenitors in snapshots between $z=7$ and 1. 
At each snapshot, 
we determine the shapes by fitting ellipsoids to sets of particles having (nearly) the same nearest neighbours-based density \citep{Jing02}.
we first compute the stellar density at the position of each stellar particle using a 80-particles kernel. The density at a given point is computed as the mass of the 80 closest particles divided by the volume of a sphere with a radius equal to the distance to the 80th 
nearest neighbour.

We compute a first guess of the major axis direction at a radius $r$ in the following way.
We define a spherical shell at that radius with a thickness of $\delta r=0.01 \kpc$. 
We compute the three eigenvectors of the inertia tensor of all stars within that shell.
The eigenvector corresponding to the smallest eigenvalue defines the first guess of the direction of the major axis.

We define a 3D isodensity surface that intersects the direction of the major axis at a radius r.
The surface is selected by a volumetric stellar density ($\rho_s$), defined as the average of the densities of the two points, located along each side of the major-axis at the intersection with the sphere of radius r.  We also compute its standard deviation ($\sigma_s$).
The isodensity surface is traced by all particles with densities around the selected value, ($\rho=\rho_s \pm \sigma_s$). In most cases, $\sigma_s < \rho_s$ and the density declines so fast with radius that the thickness of the surface is small, and we find that it is well fitted by a triaxial ellipsoidal shell.

Finally, an ellipsoidal fit to that isodensity surface defines the 3 main axes of the ellipsoidal shape.
The fit is done by diagonalizing the inertia tensor (normalized to unity mass) defined by all particles tracing the isodensity surface.
The eigenvalues ($e_1 <e_2 < e_3$) define the 3 main axes of a triaxial ellipsoidal shell \citep[][chapter 1]{Routh} as
\bea
a &=& (1.5 (e_3+e_2-e_1))^{0.5} \\
b &=& (1.5 (e_3+e_1-e_2))^{0.5} \\
c &=& (1.5 (e_2+e_1-e_3))^{0.5}
\eea
where $a$ is the major axis, $b$ is the intermediate axis, and $c$ is the minor axis of the ellipsoid. 
A new 3D isodensity surface is defined along the direction of the current major axis at a radius $r$.
The process repeats until convergence is reached, usually after 2-3 iterations, 
when the values of the principal axes do not change significantly.
The final results are profiles of the stellar shape at different radii.
A global shape is defined as the average over all shells around the stellar half-mass radius, $r=\re \pm 0.1 \ \re$,
where the half-mass radius, $\re$, is defined using all stars within 0.1$r_{\rm v}$.

The same procedure is used for dark matter. The average shape of the DM halo is taken at the same  stellar half-mass radius. Therefore, the intrinsic shape of stars and dark matter are computed at the same galactocentric radius.
The triaxiality parameter \citep{Franx91} is defined as:
\be
\rm{Tri}= \frac{a^2-b^2}{a^2-c^2}.
\label{eq:Tri}
\ee
The usual convention defines a shape as oblate if $0 \le \T < 1/3$. A prolate shape is defined by a high value, $2/3 < \T \le 1$, while triaxial shapes have intermediate values, 
and triaxiality is not defined for a sphere.

\begin{figure*} 
\includegraphics[width =0.99 \textwidth]{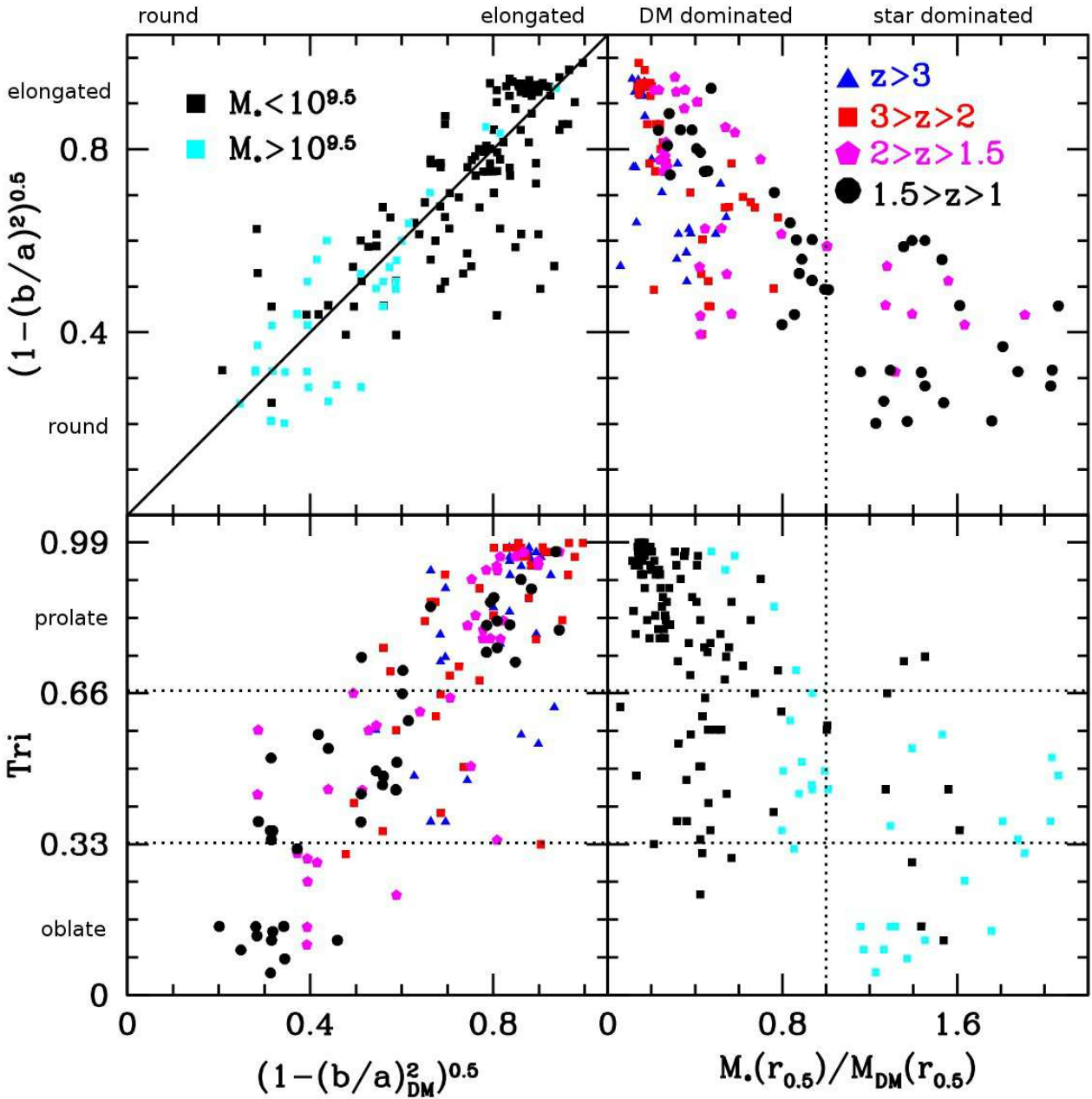}
\caption{Correlations between the elongation parameter, $e$, for stars and dark matter, the triaxiality parameter, $\T$, and the ratio of stars and DM masses at $\re$. There is a one-to-one equivalence between the elongation parameter for stars and dark matter, so that there is a large population of elongated, prolate galaxies ($\T>2/3$) especially among low-mass galaxies, $\Ms \le 10^{9.5} \ \msun$, dominated by dark matter $\Ms(\re) < {\rm M}_{\rm DM}(\re)$.
}
\label{fig:panel}
\end{figure*}

The top-left panel of  \Fig{panel} shows a one-to-one equivalence between the elongation parameter for stars and dark matter:
\be
e=\sqrt{ 1 - (b/a)^2}.
\label{eq:e}
\ee
This correspondence confirms the impression of \Fig{Vela28}. At the galaxy half-mass radius, stars and dark matter share the same intermediate-to-major axis ratio. 
We find that if
the DM shape is strongly elongated, high $e$ and low $b/a$, stars are also elongated in the same direction (as in \Fig{Vela28}). 
At $3 \le z \le 1$, the simulated galaxies of low masses, log($\Ms/\Msun) \le 9.5$, show
 a large fraction of prolate galaxies, $\T>2/3$, with a high triaxiality value of $\T=0.8 \pm 0.2$, 
a high elongation parameter, $e=0.8 \pm 0.1$, and
a relatively low intermediate-to-major axis ratio, $b/a=0.6 \pm 0.2$.
In these low-mass, prolate galaxies, dark matter dominates the mass (and the gravitational potential) inside the galaxy half-mass radius. The median ratio between  stars and dark matter masses inside $\re$ is $0.3 \pm 0.3$.
Thus, the formation of an inner elongated DM halo in the early Universe, as predicted by $\Lambda$CDM, gives rise to  galaxies elongated in the same direction.

\section{The end of elongation}
\label{sec:fall}

\Fig{panel} also shows the transition from low-mass prolate galaxies to more massive less elongated galaxies at lower redshifts, $1 \le z \le 2$.
Galaxies 
with $\Ms > 10^{9.5} \ \msun$ have 
a much lower elongation parameter, $e=0.4 \pm 0.2$,
which corresponds to
a high axis ratio, $b/a=0.9 \pm 0.1$.
Therefore, these more massive galaxies are close to axisymmetry  around the minor axis.

The key variable controlling this lack of symmetry seems to be the stellar-over-DM mass ratio inside $\re$.
The gas mass within $\re$ is lower or similar to the stellar mass.
The median axis ratio for dark-matter dominated galaxies is low, $b/a=0.6 \pm 0.2$, similar to the values of low-mass and prolate galaxies. 
The galaxies dominated by stars, $\Ms(\re) > {\rm M}_{\rm DM}(\re)$, have $b/a=0.9 \pm 0.1$, similar to the values of higher-mass, near axisymmetric galaxies.
As galaxies grow in mass, the baryons concentrate at the centres of their halos, making stars and dark matter rounder and more axisymmetric around the minor axis \citep{Zemp12}, which usually coincides with the direction of the galaxy angular momentum \citep[][and references therein]{Ceverino12,Ceverino15}.

\section{Projected axis ratios}
\label{sec:compObs}

\begin{figure} 
\includegraphics[width =0.49 \textwidth]{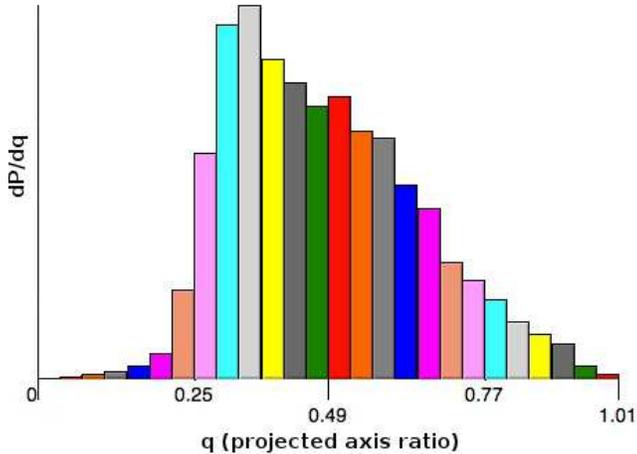}
\caption{Distribution of projected axis ratio of Vela04, using 10000 different lines-of-sight from 10 snapshots between $z=2.5$ and $z=1.5$. The  asymmetric distribution, biased towards low axis ratios, is typical of intrinsically elongated galaxies.}
\label{fig:q}
\end{figure}

Our current sample of simulations is too small for a direct and quantitative comparison with the distribution of projected axis ratios in current galaxy surveys.
However, we can make a qualitative comparison and check whether elongated galaxies can generate asymmetric distributions of projected axis ratios, as observed in current samples of low-mass galaxies at 
redshifts $0<z<2.5$ \citep{vdWel14}.

In this exercise, we take one of the simulations in our sample, Vela04, which shows an elongated morphology, similar to the example in \Fig{Vela28}.
Between $z=2.5$ and 2, it has an elongation parameter of $e\simeq0.9$ ($b/a=0.4$). At later times, the galaxy is less elongated, $e\simeq0.6$ ($b/a=0.8$) because stars dominate the mass within the half-mass radius, $\Ms(\re)/{\rm M}_{\rm DM}(\re)\simeq2$, as opposed to $\Ms(\re)/{\rm M}_{\rm DM}(\re)\simeq0.4$ at earlier times ($2.5<z<2$).
We select 10 snapshots between $z=2.5$ and $z=1.5$ and measure the projected axis ratio in 1000 random lines-of-sight per snapshot. 
The projected axis ratio is measured by fitting a 2D isocontour of surface density at $\re$ with an ellipse of axis ratio $q$.

The distribution of projected axis ratios (\Fig{q}) is clearly asymmetric. 
It is biased towards low projected axis ratios, with an asymmetric peak around $q\simeq0.35$, 
a sharp drop below $q=0.3$,
  and a long tail towards $q=1$.
This shape is qualitatively similar to the distribution shown in 3D-HST+CANDELS data for log($\Ms/ \Msun)=9-10$ at $1.5<z<2.0$ \citep[Figure 1 in][]{vdWel14}.
This comparison supports the idea that  the asymmetry in the distribution of projected axis ratios is an indication of a large population of intrinsically elongated galaxies among low-mass galaxies at high redshifts.

\section{Discussion and Conclusions}
\label{sec:conclusion}

We addressed the formation of elongated galaxies in the early Universe using five zoom-in AMR cosmological simulations of low-mass galaxies, $10^{8.8} \ \Msun \le \Ms \le 10^{9.3} \ \Msun$ at $z=2$. 
Elongated galaxies form preferentially in highly elongated, low-mass haloes at high redshifts.
This is due to asymmetric accretion from narrow filaments and mergers.
At the same time, feedback prevents the overabundance of baryons at the centre of these small haloes.
As a cautionary note, different feedback models may produce different galaxy shapes. For example, 
simulations without radiative feedback overproduce stars by a larger factor 
\citep{Ceverino14,Moody14} and generate much less prolate galaxies, according to their H-band mock images (Tang, Primack, et al. in preparation) and 3D shapes (Tomassetti, Sai,  et al. in preparation).
The gravitational potential is 
dominated by the elongated dark matter, so that the potential is highly prolate.
This is consistent with
 prolate (Tri $\simeq0.8$) galaxies, elongated in the direction of the DM major axis ($e=0.8$ or $b/a\simeq0.6$).

As galaxies grow in mass, baryons start to dominate the central potential. This makes the inner halo rounder, probably due to the deflection of elongated orbits by a central mass concentration \citep[e.g.][]{Debattista08}.
We found a decline in the fraction of elongated galaxies when stars dominate the mass inside the half-mass radius.
These galaxies are near axisymmetry and oblate ($b/a\simeq0.9$).
These massive galaxies build a disc+spheroid embedded in a rounder halo, as described in \cite{CDB, Ceverino15} and \cite{Zolotov}.

The process of wet  compaction \citep{DekelBurkert, Zolotov}, in which a gas-rich galaxy shrinks into a compact star-forming spheroid, could drive the transition from a DM-dominated elongated galaxy to a baryonic self-gravitating spheroid.
Mergers, counter-rotating streams, and the associated violent disc instability
drive baryons to the galaxy centre, increasing the central
stellar-to-DM mass ratio, making DM and baryons rounder.
The transition from DM to baryon dominance roughly occurs at the point of peak compaction
 \citep[Figs. 2 and 3 of][]{Zolotov}.

Although the stellar distribution is elongated, gas is distributed in a turbulent and thick rotating disk.
These different shapes may be related to the different angular momentum content of gas and stars.
According to the framework outlined in \cite{Danovich15}, accreted gas has a high spin, mostly coming from angular momentum acquired in the cosmic  web.
In the inner halo, gas loses some of this angular momentum due to internal torques. 
Moreover, if the inner halo is strongly elongated, the transfer of angular momentum from baryons to dark matter is  intense such that stars may lose almost all their spin and rotational support, because they spend more time in the inner halo than the freshly accreted gas. This could explain why gas forms a rotationally supported disc, while stars have an elongated, non-rotating shapes.
At later times, in more massive galaxies, their inner haloes are rounder, minimizing the angular momentum transfer and
enabling stellar discs, as shown in \cite{Ceverino15} and in \cite{Snyder}.

Intrinsically elongated galaxies may
generate a very asymmetric 
distribution of projected axis ratios, as observed in low-mass galaxies at $z \ge 1$ \citep{vdWel14}. 
Lower-mass galaxies are always more abundant than high-mass galaxies at any redshift.  
This implies that the majority of the galaxies at these redshifts are not discs or spheroids, but low-mass elongated galaxies.
If such galaxies contribute decisively to the reionization of the Universe \citep[][and references therein]{Iliev12}, 
this 
elongated morphology may have important implications for reionization.
The ability of ionizing photons to escape their host galaxies may depend on the actual galaxy geometry.
The geometry of galactic outflows may also depend on the degree of elongation of its host galaxy, because wind material  tend to escape along the direction with less resistance. 
Therefore, very elongated galaxies may host outflows with cylindrical or toroidal symmetry,
as opposed to the conical outflows expected in large discs.

\section*{Acknowledgments} 
 
We acknowledge stimulating discussions with Arjen van der Wel, Sandra Faber, David Koo, Matteo Tomassetti, and Nir Mandelker, 
and help in the analysis by Miguel Rocha and Tanmayi Sai.
The simulations were performed 
at the National Energy Research Scientific Computing Center (NERSC) at  
Lawrence Berkeley National Laboratory, and 
at NASA Advanced Supercomputing (NAS) at NASA Ames Research Center.
This work was partly supported, 
by MINECO grant AYA2012-32295,
by ISF grant 24/12,
by NSF grants AST-1010033 and AST-1405962,
and by the I-CORE Program of the PBC and The ISF grant 1829/12.

\bibliographystyle{mn2e}
\bibliography{elongated4}

\bsp

\label{lastpage}

\end{document}